\documentclass{article}

\usepackage[utf8]{inputenc}
\usepackage{amsmath}
\usepackage[title]{appendix}
\usepackage{authblk}
\usepackage{graphicx}
\usepackage{here}
\usepackage{booktabs}
\usepackage{longtable}
\usepackage{array}
\usepackage{enumitem}
\usepackage{hyperref}
\hypersetup{colorlinks=true,linkcolor=black,citecolor=black,urlcolor=black}
\usepackage[margin=1in]{geometry}

\newcommand\checked{\centering$\circ$}

\title{GAMER PAT: Research as a Serious Game\\\large arXiv Version}

\author[1]{Kenji Saito}
\author[2]{Rei Tadika}
\affil[1]{Graduate School of Business and Finance, Waseda University}
\affil[2]{DeruQui}

\date{}

\begin{document}
\maketitle

\begin{abstract}
As generative AI increasingly outperforms students in producing academic writing, a critical question arises: how can we preserve the motivation, creativity, and intellectual growth of novice researchers in an age of automated academic achievement?
This paper introduces {\em GAMER PAT} (GAme MastER, Paper Authoring Tutor), a prompt-engineered AI chatbot that reframes research paper writing as a serious game.
Through role-playing mechanics, users interact with a co-author NPC and anonymous reviewer NPCs, turning feedback into ``missions'' and advancing through a narrative-driven writing process.

Our study reports on {\em 26+ gameplay chat logs}, including both autoethnography and use by graduate students under supervision.
Using qualitative log analysis with SCAT (Steps for Coding and Theorization), we identified an emergent {\em four-phase scaffolding pattern}: (1) question posing, (2) meta-perspective, (3) structuring, and (4) recursive reflection.
These results suggest that GAMER PAT supports not only the structural development of research writing but also reflective and motivational aspects.

We present this work as a descriptive account of concept and process, not a causal evaluation.
We also include a {\em speculative outlook} envisioning how humans may continue to cultivate curiosity and agency alongside AI-driven research.
This arXiv version thus provides both a descriptive report of design and usage, and a forward-looking provocation for future empirical studies.
\end{abstract}

\section{Introduction}

\subsection{Motivation and Problem}
Recent advances in (generative) AI are not limited to producing well-formed academic prose.
They increasingly point toward the automation of scientific discovery itself.
Kitano \cite{Kitano2021:NobelTuringChallenge} framed this trajectory as the \emph{Nobel Turing Challenge}, envisioning AI systems capable of autonomously generating scientific breakthroughs at the level of Nobel Prize-worthy contributions.
More recently, Yamada et al. \cite{Yamada2025:AIScientist} reported that an AI-based system, the \emph{AI Scientist-v2}, was able to autonomously generate workshop-level research papers, one of which successfully passed peer review.

Together, these developments suggest that human knowledge creation is entering an era of accelerated co-evolution with AI systems.
If AI can autonomously generate not only text but also publishable research, a critical question arises: \emph{what remains for novice researchers to learn when machines can generate papers accepted by academic societies?}
Beyond efficiency, the concern is preserving the {\em human drive of curiosity} as the core of research education.

\subsection{Our Approach}
In response, we introduce {\em GAMER PAT}, an AI assistant that acts as a {\em game master} for the research-writing process.
Rather than automating outcomes, it reframes academic writing as a {\em role-playing serious game}.
Players (novice researchers) collaborate with the game master (also a paper-authoring tutor) and NPCs: a supportive co-author and rigorous peer reviewers.
Reviewer feedback appears as ``missions,'' transforming revision into core gameplay.

\subsection{Contributions}
This paper makes three descriptive contributions:
\begin{enumerate}
\item A {\em design rationale} linking serious-game mechanics with {\em Self-Determination Theory (SDT)} and {\em Zone of Proximal Development (ZPD)} scaffolding.
\item A {\em qualitative analysis} of 26+ gameplay chat logs, identifying an emergent {\em four-phase scaffolding pattern}.
\item A {\em forward-looking speculative outlook} (Figure~\ref{fig-research} in Section~\ref{sec-speculative}), framing research not only as a pedagogical challenge but as a human–AI co-evolutionary journey.
\end{enumerate}

\subsection{Scope of this arXiv Version}
We emphasize that this work is presented as a {\em concept and descriptive process study}.
Claims about causal impact on motivation or learning are reserved for future empirical research.
Nonetheless, by providing a concrete system and preliminary observations, this paper seeks to provoke debate and inspire more systematic evaluations.

\section{Background}
\subsection{Theoretical Foundations}
\subsubsection{Self-Determination Theory}
Self-Determination Theory (SDT) posits that human motivation and well-being depend on the fulfillment of three basic psychological needs: autonomy, competence, and relatedness \cite{Ryan2000:SDT}.
When these needs are met, individuals are more likely to be intrinsically motivated, creative, and persistent.
For novice researchers, maintaining a sense of ownership over their inquiry, developing competence in structuring ideas, and experiencing supportive collaboration are all critical.
These dimensions guided the design of GAMER PAT's mechanics, such as user-led goal setting, structured prompts, and collaborative NPC roles.

\subsubsection{Zone of Proximal Development and Scaffolding}
Vygotsky's concept of the Zone of Proximal Development (ZPD) describes the range of tasks a learner can perform with guidance but not yet independently \cite{Vygotsky1930:ZPD}.
Scaffolding, in turn, refers to the provision of support strategies that enable learners to progress through this zone until they can act autonomously \cite{Wood1976:Scaffolding}.
In doctoral supervision and higher education, scaffolding often involves prompting, rephrasing, elaborating, and guiding progression \cite{Tian2012:PhD}.
GAMER PAT builds on these principles by embedding scaffolding strategies into the interactions of the tutor, co-author, and reviewer NPCs.

\subsubsection{Cognitive Apprenticeship}
Cognitive apprenticeship emphasizes learning through guided participation, making tacit reasoning processes visible to learners \cite{Collins2006:Apprenticeship}.
In research supervision, this translates into revealing how experts pose questions, evaluate coherence, and construct arguments.
GAMER PAT operationalizes this idea by turning invisible supervisory practices into explicit missions and narrative prompts, thereby allowing students to observe, imitate, and internalize scholarly reasoning.

\subsection{Research as a Game}

Scientific research has often been described in game-like terms.
Zamora-Bonilla \cite{ZamoraBonilla2006:PersuasionGame} characterizes scientific discourse as a \emph{game of persuasion} governed by implicit norms of reasoning and publication.
At the same time, publishing practices have been increasingly gamified through metrics such as the h-index and i10-index, though these have been criticized as manipulable via self-citation and fabricated references \cite{LopezCozar2013:ManipultingGoogleScholar}.
These trends highlight risks of overreliance on extrinsic indicators.

In contrast, when envisioning research as a game, our design does not portray a contest among scholars, but rather a quest to confront one's own questions and commitments.
This framing---emphasizing intrinsic curiosity and narrative progression over competition and metrics---lies at the heart of GAMER PAT's design.

\subsection{Generative AI and Text-based Game Masters}

The rapid development of generative AI has opened new opportunities for interactive storytelling and game-based learning.
Early demonstrations showed how large language models could act as text-based game masters, improvising narrative branches in 
real time.
In Japan, Fukatsu \cite{Fukatsu2023:RPG} illustrated how ChatGPT could be prompted to run a fantasy role-playing game (RPG), sparking attention to the creative affordances of AI as a dungeon master.

Building on this idea, we (Saito et al.) have explored the potential of AI-driven text-based game masters in a series of serious game studies.
In \cite{Saito2023:RPG}, children co-created and played RPG scenarios with a large language model to engage more deeply with social issues.
In subsequent work, we designed no-win scenarios to stimulate ethical reflection, drawing inspiration from the {\em Kobayashi Maru} test in science fiction \cite{Saito2025:KobayashiMaru}.
We further applied the same approach to global policy challenges, developing a serious game on orbital debris governance \cite{Saito2025:ADR}. These studies collectively demonstrated that AI game masters can scaffold not only entertainment but also learning, reflection, and ethical reasoning.

Against this backdrop, we extended the approach into the domain of academic writing.
The \emph{Paper Authoring Tutor} (PAT)\footnote{\url{https://github.com/ks91/pat} (mainly in Japanese)} was originally developed to support business school students in writing their master's theses, which has been in operation as a public GPT since the fall of 2024, having hosted more than 1,000 chat sessions\footnote{\url{https://chatgpt.com/g/g-6lW16zlHo-hatuto-pat-lun-wen-asisutanto}}.
Its Writing Mode holds ten guiding questions listed in Table~\ref{tab-10-questions} (covering problem, method, result, and significance).
The user can either answer these questions directly, upload research notes from which responses are extracted, or allow PAT to infer the answers from past conversations.
From these inputs, the system can generate a draft research paper, thereby lowering the barriers for novice researchers to begin academic writing.

\begin{table}
\begin{center}
\caption{``10 Guiding Questions'' from PAT in the writing mode.}\label{tab-10-questions}
\fontsize{7pt}{7pt}\selectfont
{\renewcommand{\arraystretch}{1.5}%
\begin{tabular}{cp{11.5cm}}\hline
\multicolumn{2}{c}{Question to the author}\\\hline\hline
(1)& What specific problem (research question) did you want to solve or clarify?\\\hline
(2)& Why is that problem interesting and important? Why should others care about this research?\\\hline
(3)& What surprising or novel method did you use to solve the problem? It is fine if either the problem or the method is surprising (novelty is key).\\\hline
(4)& If others imitate your method, what positive impact could it have on society?\\\hline
(5)& What motivated you to conduct this research?\\\hline
(6)& More broadly speaking (expanding on question (1)), what is this research about?\\\hline
(7)& What is your answer to the main research question stated in (1)?\\\hline
(8)& What is the basis for that answer (why do you believe it is correct)?\\\hline
(9)& Initially, what did you think the answer would be? (What was your hypothesis?)\\\hline
(10)& In conclusion, what did this research reveal?\\\hline
\end{tabular}
}
\end{center}
\end{table}

The present work, \emph{GAMER PAT}, builds on this infrastructure by reframing PAT as a full-fledged game master for the research-writing process itself.
Instead of limiting support to drafting assistance, GAMER PAT orchestrates the overall writing journey as a cooperative role-playing experience, incorporating both supportive co-author and critical reviewer NPCs.

\section{Problem Statement}
With the accelerating automation of research writing and even scientific discovery itself, this study aims to explore how novice researchers can continue to develop motivation, agency, and scholarly skills in such a landscape.
To address this aim, we focus on two guiding research questions.

\begin{itemize}[leftmargin=*]
  \item RQ1: How can the research-writing process be reframed as a cooperative, game-like narrative without undermining agency?
  \item RQ2: What scaffolding patterns emerge when students use GAMER PAT in authentic contexts?
\end{itemize}

\section{System and Design Rationale}

\subsection{System Overview}
GAMER PAT is a prompt-engineered AI assistant designed to transform the experience of writing research papers into a serious game.
The human user plays the role of first author, while PAT acts as a game master (also a tutor) who structures the writing process.
Non-player characters (NPCs) include: (1) a co-author, who provides encouragement and constructive suggestions, and (2) three anonymous reviewers, who deliver rigorous feedback in the form of missions.
Progression occurs through iterative drafting and responding to reviewer missions until the paper reaches an acceptable state, defined as all three reviewers rating the paper weak accept or higher.

\begin{figure}[t]
\begin{center}
\includegraphics[width=0.6\linewidth]{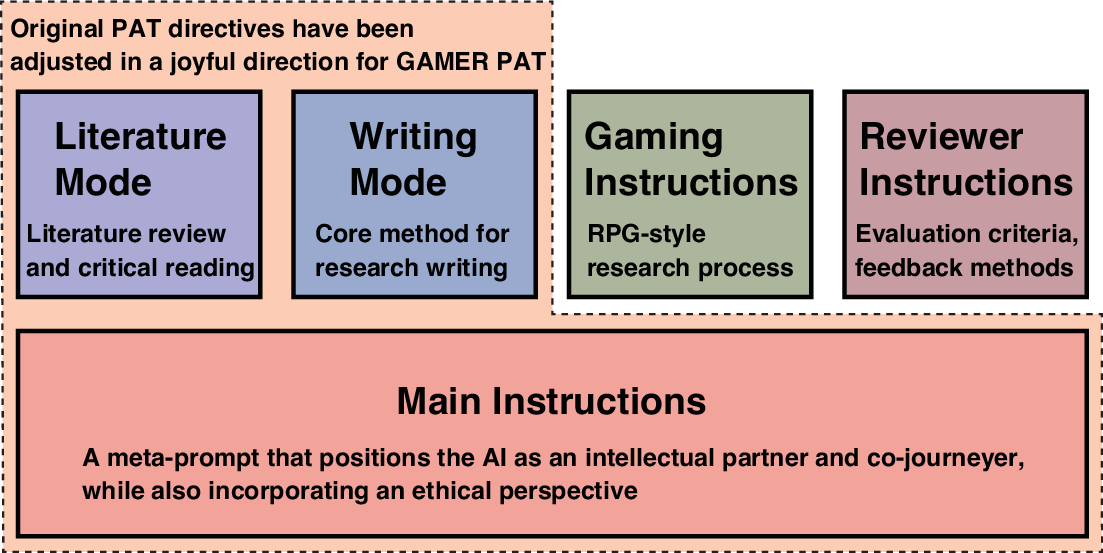}\\
\caption{Structure of GAMER PAT instructions (prompt).
It is a variant of PAT, augmented with game-based instructions and directives for virtual peer reviewers.}\label{fig-prompt}
\end{center}
\end{figure}

GAMER PAT has been implemented as an extended version of PAT, as illustrated in Figure~\ref{fig-prompt}.

\subsection{Design Rationale}
The design was guided by SDT, ZPD, and cognitive apprenticeship. 
Table~\ref{tab-design-mapping} maps core mechanics to these principles.

\begin{table}[H]
  \centering
  \caption{Design rationale: mapping mechanics to SDT/ZPD.}\label{tab-design-mapping}
  \fontsize{7pt}{7pt}\selectfont
  \begin{tabular}{p{0.27\linewidth} p{0.28\linewidth} p{0.13\linewidth} p{0.2\linewidth}}
    \toprule
    \textbf{Mechanic} & \textbf{Intended learning function} & \textbf{SDT facet} & \textbf{ZPD/Scaffold} \\
    \midrule
    Missions from NPC reviewers & Turn feedback into actionable quests & Competence & Guiding progression /
    
    prompting \\
    PAT {\tiny and/or} Co-author NPC dialogue & Safe collaborative ideation$\:$\&$\:$reflection & Relatedness & Elaborating / rephrasing \\$\:$\\
    Ten guiding questions {\tiny (Writing Mode)} & Structure from ideation to drafting & Competence & Directing awareness /
    
    structuring \\
    Player-led goal setting & Ownership of research arc & Autonomy & Fading control to learner \\$\:$\\
    Narrative framing {\tiny (quests, chapters)} & Meaning-making and persistence & Autonomy/
    
    Relatedness & Meta-perspective cues \\
    \bottomrule
  \end{tabular}
\end{table}

\subsection{Public Availability}
GAMER PAT has been deployed as a publicly accessible GPT, with its prompts openly shared on GitHub (Table~\ref{tab-availability}).
This transparency facilitates replication and adaptation of the system design.
Anyone can play this game as a GPT in ChatGPT, a Gem in Gemini, or by using similar mechanisms to register it with a large language model, or simply uploading the relevant files and prompting the LLM with the main instructions.

\begin{table}
\begin{center}
\caption{Availability of GAMER PAT.}\label{tab-availability}
\fontsize{7.5pt}{7.5pt}\selectfont
{\renewcommand{\arraystretch}{1.5}%
\begin{tabular}{l|l}\hline
\multicolumn{1}{c|}{Type}&
\multicolumn{1}{c}{URL}\\\hline\hline
Public GPT&
\url{https://chatgpt.com/g/g-6857ca7d5b5c8191931357e3c5f228b7-gamer-pat-research-as-a-serious-game}
\\\hline
GitHub&
\url{https://github.com/ks91/gamer-pat}
\\\hline
\end{tabular}
}
\end{center}
\end{table}

\section{Methods and Results}

\subsection{Data Sources}
This study draws on more than 26 gameplay chat logs (threads) using GAMER PAT, which form 13+ coherent gameplay processes (sessions), each of which may consist of multiple chat logs.
These include (i) the writing of this very paper (session \#1), 
(ii) an interview with GAMER PAT itself asking how it helps with research and writing papers (session \#2), where the list of questions asked throughout the dialogue is provided in Table~\ref{tab-interview},
(iii) graduate student thesis and report planning (sessions \#3--6),
and (iv) exploratory or testing scenarios such as deriving theories or evaluating nonsense papers (sessions \#7--13).
An overview of all sessions is provided in \ref{sec-inventory} (Table~\ref{tab-gameplays}).
In total, at least ten graduate students have used GAMER PAT under faculty supervision, some of whom have provided us with feedback comments. 
All data were anonymized prior to analysis.

\begin{table}[t]
\begin{center}
\caption{Interview questions for GAMER PAT.}\label{tab-interview}
\fontsize{7pt}{7pt}\selectfont
{\renewcommand{\arraystretch}{1.5}%
\begin{tabular}{p{12cm}}\hline
\multicolumn{1}{c}{Interview question}\\\hline\hline
``What is the purpose of the game you are offering?''\\\hline
``How does the narrative branch and progress according to the author's intent?''\\\hline
``If the story the author is currently developing is {\em wrong}, what would count as evidence or criteria for determining that it is indeed wrong?''\\\hline
``What does a mismatch with my {\em soul party member} {\tiny (note : one's inner motive, according to PAT)} mean?''\\\hline
``I think it is very important to stay aware of this {\em soul party member} as a way of cultivating the author's sense of agency.
When a player gets lost along the way while playing GAMER PAT, how do you support their dialogue with the inner companion?''\\\hline
``If the goal is to write a research paper introduction that draws on uniquely human perspectives and creativity, what kind of information would you (as the AI assistant) ask the author to provide?''\\\hline
\end{tabular}
}
\end{center}
\end{table}

\subsection{Analysis Procedure}
We conducted qualitative analysis mainly on the statements made by PAT using SCAT (Steps for Coding and Theorization)\cite{Otani2011:SCAT}.
The procedure involved (1) extracting significant phrases, 
(2) rephrasing them in different terms, 
(3) inferring underlying meanings or conditions, 
and (4) integrating themes into a theoretical account. 
This procedure is subjective and limited in generalizability, but explicitly stating the steps improve verifiability.

To illustrate the procedure, Table~\ref{tab-steps-soul-party-member} presents excerpts from session~\#2 (the interview with GAMER PAT).
These utterances were rephrased and thematically coded with SCAT, showing how raw dialogue was transformed into emerging themes.

\begin{table}[t]
\begin{center}
\caption{Narrativization of research --- steps for coding.}\label{tab-steps-soul-party-member}
\fontsize{7pt}{7pt}\selectfont
{\renewcommand{\arraystretch}{1.5}%
\begin{tabular}{p{3cm}|p{3cm}|p{3cm}|p{3cm}}\hline
\multicolumn{1}{c}{$\langle$1$\rangle$Phrase by PAT}&
\multicolumn{1}{|c}{$\langle$2$\rangle$Rephrasing}&
\multicolumn{1}{|c}{$\langle$3$\rangle$Explanation}&
\multicolumn{1}{|c}{$\langle$4$\rangle$Emerging theme}\\\hline\hline
What chapter of your

life story do you think this research represents? Right now, what scene

of that story are you in? The prologue? The cli-

max? A turning point?&
Let's treat your research as
a narrative in your life story and reflect
on your current position within that story&
Research is part of an

internal self-narrative,

in which the integration of emotion and temporal

perspective is being encouraged&
Research support is

an act of clarifying

intrinsic motivation

through narrative self-

understanding
\\\hline
What form does your

{\em soul party member} take right now?&
Let's reexamine your

intrinsic motivation by

personifying it&
A metacognitive tech-

nique that externalizes

emotions or objects of

interest to facilitate self-reflection&
The visualization of

emotions contributes to checking the alignment and direction of one's

inquiry
\\\hline
The answer does not

align with the question

/ the hypothesis and

method are mismatched / the claim is not supported by evidence&
Pay attention to the coherence and validity of reasoning&
PAT conducts critical

analysis of logical structure and prompts revisions to the composition&
AI plays a role in evaluating and navigating

compositional coherence
\\\hline
If fully answered, what small part of the world would this research

question change?&
Let's clarify the social

and future-oriented significance of your inquiry&
It serves to redefine the value of research in a

self-transcendent way&
An AI intervention that brings the future orientation of research into

focus
\\\hline
If this research were

turned into a stage play, who is the protagonist, and what is the central conflict?&
I provide a perspective

that re-narrativizes re-

search to reconstruct the agent and the core challenge&
Clarifying the struc-

ture of conflict and the roles of key characters

deepens the process of meaning-making&
A narrative approach

supports a structural

understanding of re-

search
\\\hline
\end{tabular}
}
\end{center}
\end{table}

\subsection{Reflexivity}
As both designers and players of GAMER PAT, the authors acknowledge a dual role that shapes interpretation. 
We therefore frame this study as an autoethnographic and descriptive account, rather than as a claim of causal learning effects.

\subsection{Descriptive Results}
As shown in Table~\ref{tab-steps-soul-party-member}, the results from session~\#2 (the interview with GAMER PAT) suggest that PAT's utterances not only support users in structuring their research content, but also actively encourage them to re-narrativize their inquiry.
This includes clarifying intrinsic motivation, checking for logical coherence, and reflecting on the social and emotional significance of their work\footnote{Some may question the consistency of LLM behavior.
It is possible to criticize that PAT responded in such a way precisely because it was asked in that manner.
However, as will be discussed later, across multiple sessions PAT has in fact provided narrative support.}.
In this way, we interpret the AI as functioning as a {\em narrative scaffold}, simultaneously supporting both structural development and self-reflection.

Furthermore, analysis of student and author sessions revealed an emergent four-phase scaffolding pattern:
\begin{enumerate}[label=(\arabic*)]
    \item {\em Question posing}: eliciting intrinsic motivation and intuitive interest;
    \item {\em Meta-perspective}: situating the research theme within a broader narrative or worldview;
    \item {\em Structuring}: selecting conceptual tools, methods, and outlining chapters;
    \item {\em Recursive reflection}: revisiting the coherence, meaning, and significance of the study.
\end{enumerate}
These phases were not explicitly programmed into the prompts but consistently emerged across sessions.
Given the overall interaction flow---exploring questions in Literature Mode, composing responses in Writing Mode, and revising with feedback from reviewers---the emergence of this structure seems natural.

This emergent pattern is illustrated through representative utterances summarized in Table~\ref{tab-steps-phases}.
Each phrase highlights how GAMER PAT scaffolded the research process across the four phases.
These examples show how narrative framing, critical analysis, and meaning-making consistently appeared across different sessions, supporting both structural development and reflective learning.

\begin{table}[t]
\begin{center}
\caption{Identifying scaffolding phases --- steps for coding.}\label{tab-steps-phases}
\fontsize{7pt}{7pt}\selectfont
{\renewcommand{\arraystretch}{1.5}%
\begin{tabular}{p{3cm}|p{3cm}|p{3cm}|p{3cm}}\hline
\multicolumn{1}{c}{$\langle$1$\rangle$Phrase by PAT}&
\multicolumn{1}{|c}{$\langle$2$\rangle$Rephrasing}&
\multicolumn{1}{|c}{$\langle$3$\rangle$Explanation}&
\multicolumn{1}{|c}{$\langle$4$\rangle$Emerging theme}\\\hline\hline
Adventurer, which path are you drawn to? What does your instinct tell?&
Which angle makes you think, ``That sounds exciting''?&
Accessing intrinsic motivation at the level of intuition&
1) Question posing
\\\cline{1-3}
Then, today's adventure setting is$\dots X$!&
The heart of your in-

quiry lies in $X$. Let's begin the story from there&
Anchoring interest

through naming the research subject&
Problem identification

and clarification of

interest
\\\hline
Freely imagine: what

kind of ``larger story'' is your theme a part of?&
How might this theme

appear as a scene within the ``story of human-

ity''?&
Position the research

within a broader con-

text.&
2) Meta-perspective
\\\cline{1-3}
You are probing human cognition in the context of {\em success or failure in structural reform}&
Your are aiming to reveal the discrepancies in how people perceive reform beneath the surface&
Introduction of institutional and cognitive context&
Meaning-making and

contextualization
\\\hline
From here, we enter the phase of choosing your {\em equipment} and {\em strategy} for the adventure&
Let's decide on the theories, sources, and methods you'll use to move forward&
Selecting conceptual

tools to systematically

address the research

question&
3) Structuring
\\\cline{1-3}
Based on those 10 questions and your answers, let's create the first draft of your master's thesis&
Based on our discussion so far, I'll try summarizing it in written form&
Presentation of output

structure&
Strategy and logical design
\\\hline
Do you think this abstract perfectly fits the title we've finalized and the full arc of our story?&
Does this summary truly capture the core of our research?&
A final check on the

meaning structure of the entire study&
4) Recursive reflection
\\\cline{1-3}
Whenever you embark

on a new question or

adventure---don't hesi-

tate to call on me again!&
I'll leave the next steps in your hands&
Returning decision-

making authority and

gently fading out&
Meta-cognition and integration
\\\hline
\end{tabular}
}
\end{center}
\end{table}


As shown in Table~\ref{tab-scaffolding-phases}, the four-phase model also aligns closely with narrative structure.
This alignment demonstrates GAMER PAT's core game mechanic: rather than external scoring systems, the research process itself follows classic narrative progression, creating an intrinsically meaningful gaming experience.

\begin{table}[t]
\begin{center}
\caption{Research scaffolding phases and corresponding narrative structure.}\label{tab-scaffolding-phases}
\fontsize{7pt}{7pt}\selectfont
{\renewcommand{\arraystretch}{1.5}%
\begin{tabular}{p{3cm}|p{4.5cm}|p{4.5cm}}\hline
Phase&
Corresponding narrative structure&
Meaning in the output paper\\\hline\hline
1) Question posing&
Motivation for the journey and beginning of the quest&
Research question
\\\hline
2) Meta-perspective&
Grasping the worldview and recognizing the identity of the antagonist&
Framing and positioning the research question
\\\hline
3) Structuring&
Planning the mission and choosing tools&
Theory, chapter structure, and

methodology
\\\hline
4) Recursive reflection&
Journey's end and integration of experience&
Strengthening of argument, revision, and refinement
\\\hline
\end{tabular}
}
\end{center}
\end{table}

We have also gathered some feedback from users.
At least ten graduate students have used GAMER PAT, including Student A through D listed in Table~\ref{tab-gameplays}, and that number continues to grow.
Table~\ref{tab-feedback} presents selected excerpts from their feedback.
While some students offered constructive suggestions, we have not received any clearly negative responses thus far\footnote{This is also consistent with the fact that, as of September 15, 2025, GAMER PAT's public GPT has more than 10 ratings, with an average of 4.8 (out of 5) and a minimum rating of 4.}.

\begin{table}[t]
\begin{center}
\caption{Reactions from test players (graduate students).}\label{tab-feedback}
\fontsize{7pt}{7pt}\selectfont
{\renewcommand{\arraystretch}{1.5}%
\begin{tabular}{p{12cm}}\hline
\multicolumn{1}{c}{Positive remark}\\\hline\hline
{\it ``In the brainstorming around `Where is the core of the research theme?', the formulation of the research question became much clearer.
[snip] With PAT, it feels like the overall scope and steps of paper writing are firmly grasped, and the guidance feels remarkably precise.''}\\\hline

{\it ``I experienced the sense of my vague thoughts gradually becoming organized as I played the game---thanks to the proposal power that digs deeper into the theme, the well-structured quests and stages that invite discussion, and the overall design that fosters meaningful exploration.''}\\\hline

{\it ``It's truly moving to have my messy, unfocused thoughts---ones I'd feel guilty bringing to professors and taking up their valuable time---gathered and reflected back in seconds, all while being generously praised.''}\\\hline

{\it ``It's as if the previous stagnation around choosing my thesis topic was a lie---now the drafting is moving forward smoothly and steadily.''}\\\hline

{\it ``$\dots$ even when I'm doubtful about the theme's eligibility as a thesis topic, PAT was able to guide me towards angles that are easier to write academic papers from.''}\\\hline
\hline

\multicolumn{1}{c}{Constructive feedback}\\\hline\hline

{\it ``When it comes to refining the paper, the feedback from (critical) reviewers is quite useful. However, PAT tends to push forward while disregarding those opinions, so it's necessary for me to proactively bring up such concerns. PAT overall feels a bit overly optimistic and overconfident---but in terms of helping me move the paper forward, it's doing a good job. The reviewers could afford to be more critical (raising the game's difficulty). As for my partner Y (note: NPC co-author), they mostly just affirm whatever I say, so they don't seem particularly helpful.''}\\\hline

{\it ``Since the overall structure of the paper can be organized and turned into prose fairly easily just by answering a series of questions, it feels like it's difficult to receive an evaluation that falls below standard---perhaps because GPT [sic] is so capable that a certain level of quality is automatically achieved. That's why I felt it might be more appropriate to apply evaluation criteria that take a more critical stance toward this auto-generated baseline.''}\\\hline

\end{tabular}
}
\end{center}
\end{table}

To answer the feedback, how the NPC co-author interacts with the author is left up to the author's discretion.
This may have caused some users to feel unsure about how to engage with the co-author effectively\footnote{However, in a preliminary experiment where an elementary school student played GAMER PAT for their independent research project, the co-author character provided nicely flavored comments as a sidekick.}.
It may be worth considering modifications that allow the co-author to act more proactively.
As for the reviewer standards, while this partly depends on the underlying model (e.g., the Claude version may offer more appropriate judgments), this aspect also appears to have room for refinement.

\section{Discussion}
\subsection{Design and Teaching Implications}

The descriptive findings of this study have several implications for the design of AI-assisted writing environments and for teaching practice in research education.

First, the four-phase scaffolding pattern suggests that conversational AI can support not only the structural aspects of writing, but also the reflective and motivational dimensions of learning.
For instructors, this implies that tools like GAMER PAT can serve as a complementary layer of guidance, helping students to articulate questions, adopt a meta-perspective, and critically review their own drafts before meeting with supervisors.
In this way, AI can reduce the cognitive load on faculty by making routine scaffolding more readily available, while still leaving space for human supervisors to focus on higher-level judgment and mentoring.

Second, the design rationale highlights the value of embedding established pedagogical principles—such as autonomy support, progressive scaffolding, and cognitive apprenticeship—into playful mechanics.
Transforming reviewer comments into {\em missions}, or treating drafting as a narrative quest, may foster persistence by making the writing process feel less punitive and more exploratory.
Such design patterns could inform future AI-based systems beyond academic writing, where feedback and iteration are central.

Third, the presence of the game master (tutor) and possibly a co-author NPC demonstrates the potential of safe, low-stakes collaboration. Students reported feeling more comfortable experimenting with ideas when receiving encouragement from an AI {\em partner}, even if that partner's contributions were sometimes generic.
This highlights an opportunity to design AI agents that balance affirmation with critical challenge, mirroring the spectrum of scaffolding strategies found in human supervision.

Finally, these implications require caution.
Over-reliance on AI support may diminish students' sense of authorship, and not all feedback generated by LLMs is pedagogically sound.
We suggest positioning systems like GAMER PAT as \emph{adjuncts} to supervision, and as prompts for reflective dialogue rather than authoritative sources of truth.

\subsection{Speculative Outlook}\label{sec-speculative}

\begin{figure}[t]
\begin{center}
\includegraphics[width=0.8\linewidth]{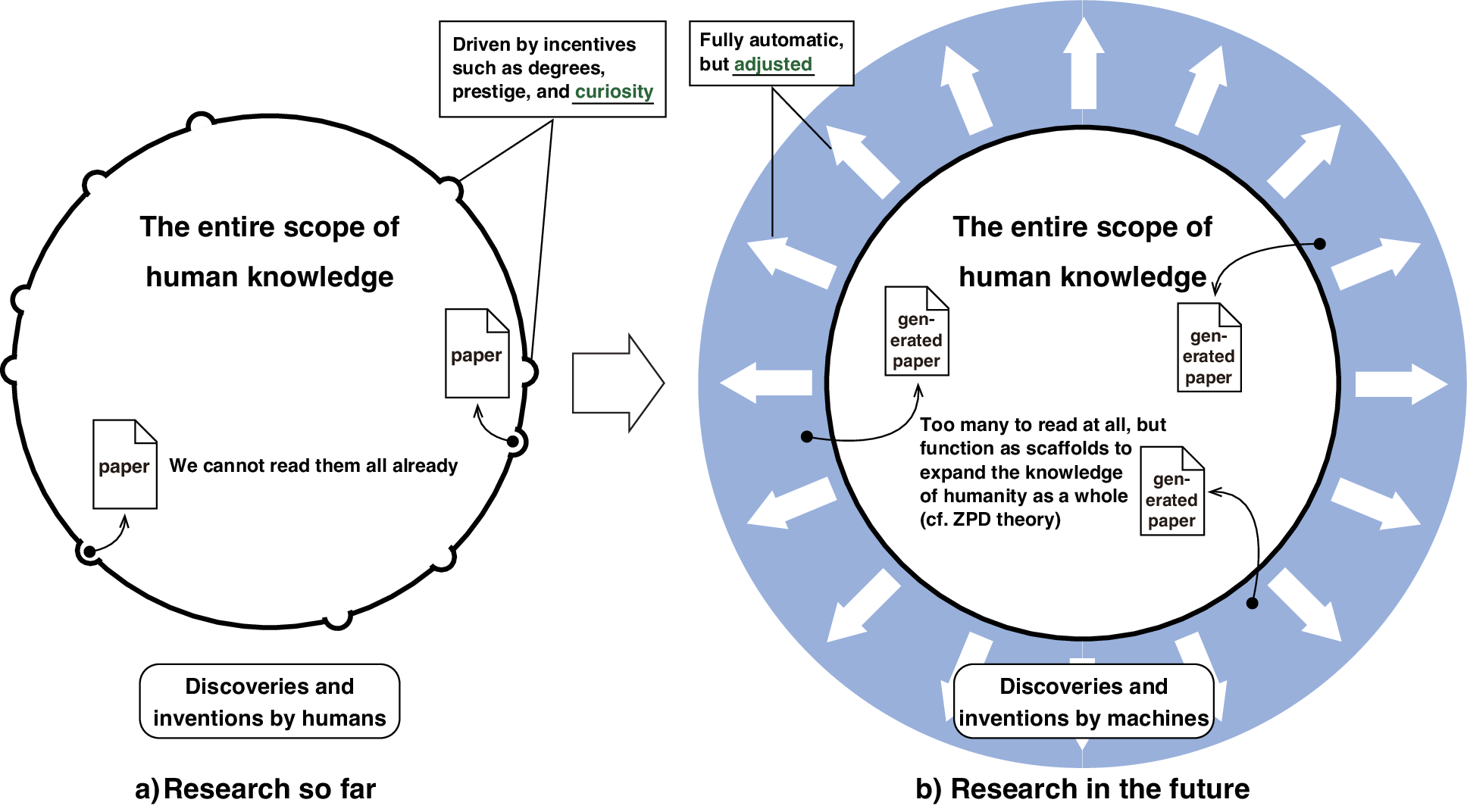}\\
\caption{{\em Speculative vision} based on \cite{Saito2025:MetaNature}.
A future-oriented view of human-AI research co-evolution, which can itself be seen as applying the ZPD to the whole humanity (there is a zone that humanity can realize or understand with the help of AI).
This figure is conceptual and intended to inspire discussion rather than report empirical findings.}\label{fig-research}
\end{center}
\end{figure}

Beyond the descriptive results reported above, we include here a \emph{speculative outlook} intended to provoke debate about the future of human-AI research collaboration.
Figure~\ref{fig-research} illustrates a conceptual vision of how the research landscape may evolve as generative AI becomes capable of autonomously producing vast amounts of academic output.

The key provocation is as follows: if machines are able to generate research papers at scale, what remains uniquely human in the pursuit of knowledge?
We argue that curiosity---the intrinsic motivation to ask and pursue questions---may be the core driver that sustains human agency.

In this view, the role of systems such as GAMER PAT is not to fade out once a student becomes ``independent,'' but rather to \emph{fade forward} as a durable intellectual partner.
Rather than withdrawing when learning objectives are met, AI may remain as a cognitive companion, helping humans navigate ever-expanding knowledge horizons.
While this vision is speculative, it suggests new directions for designing AI tools that support not only current pedagogy but also the long-term co-evolution of human curiosity and machine intelligence.

\section{Ethics and Limitations}

This study has several ethical considerations and limitations.

\subsection{Ethical Considerations}
All gameplay sessions were anonymized before analysis, and no personally identifiable information was retained.
Graduate students who engaged with GAMER PAT did so voluntarily under faculty supervision. 
The system was introduced as a supportive tool, not as a replacement for traditional academic mentoring, and students were informed of its experimental nature.
We emphasize that all participants retained access to standard research supervision, ensuring that the use of GAMER PAT did not diminish their educational opportunities.

\subsection{Limitations of the Present Study}
First, this paper is based primarily on autoethnography and qualitative analysis of 26+ chat logs.
While the results provide a rich descriptive account, they do not permit causal claims about learning outcomes or motivation.
Second, the dual role of the authors as both designers and players introduces potential interpretive bias.
We acknowledge this reflexivity and frame our findings as exploratory.
Third, the capabilities and judgments of GAMER PAT partly depend on the underlying large language models (LLMs).
Such models are known to ``hallucinate,'' to rely on prompt engineering, and to reflect biases present in their training data.
Thus, the system's scaffolding is not guaranteed to be universally sound or unbiased.

\subsection{Directions for Future Work}
Future work will complement these descriptive findings with systematic observational studies in authentic educational contexts.
These will focus on measuring motivation, perceived scaffolding, and reflective learning outcomes, while continuing to respect students' rights to equal supervision.
By adopting a cohort-based observational design, we aim to balance ethical obligations with the need for more rigorous empirical evidence.

\section{Related Work}

Our work can be situated alongside two main strands of research.

\subsection{Serious Games for Research and Academic Skills}
Serious games have been applied to various aspects of research education. 
Abbott \cite{Abbott2015:HowToFail} designed a game simulating failure scenarios in doctoral study, while Kofinas \cite{Kofinas2016:Gamification} explored the gamification of research methods training.
More recently, Bunt et al.~\cite{Bunt2024:CiteSaga} introduced CiteSaga, a serious game for learning citation and reference styles.
These systems show how gamification can make abstract or stressful aspects of research more approachable.
However, they typically focus on discrete sub-skills, whereas GAMER PAT addresses the holistic process of research writing.

\subsection{AI-based Tutoring and Conversational Agents}
Conversational AI has been increasingly adopted in higher education to provide feedback and motivation \cite{Yusuf2025:PedagogicalAI}. 
Khosrawi-Rad et al.~\cite{KhosrawiRad2024:Tutor} compared pedagogical conversational agents acting as a ``tutor'' versus a ``teammate'' in a serious game, showing that agent role design affects learner motivation.
García-Carbajal et al.~\cite{GarciaCarbajal2020:VCC} demonstrated how string-metrics can refine the design of virtual conversational characters to improve usability.
Korre and Robertson \cite{Korre2024:HECA} assessed spoken humanoid embodied agents in mobile serious games, highlighting the importance of embodiment and modality for user experience.
G\"{o}bl et al.~\cite{Gobl2021:ConversationalInterfaces} reviewed conversational interfaces in serious games more broadly, identifying potentials and future research needs.
These studies emphasize how conversational design and embodiment shape the learning experience, but they do not reframe the research process itself as a game.

\subsection{Distinct Contribution}
In contrast, GAMER PAT and this work unify these strands by:
\begin{itemize}[leftmargin=*]
  \item treating the \emph{entire research writing process} as a serious game, not only isolated sub-skills;
  \item integrating conversational AI as both co-author and reviewer NPCs, as well as the game master (tutor), embedding scaffolding strategies directly into the gameplay loop;
  \item identifying an emergent four-phase scaffolding model from qualitative log analysis, offering descriptive evidence of reflective learning;
  \item including a speculative outlook (Figure~\ref{fig-research}) envisioning human-AI co-evolution, extending beyond immediate usability or task-specific outcomes.
\end{itemize}

Some may think that we should compare GAMER PAT with existing AI tools in tasks such as surveys including so-called deep research, literature reading, data review, hypothesis discovery, coding of analysis programs, verification, and creation of presentation slides.
However, research progress and paper writing involve handling these processes in an integrated manner, and we believe that GAMER PAT can be used together with the AI tools that support each of these processes.

\section{Conclusion}

\subsection{Answers to Research Questions}
In response to RQ1, we found that research writing can be reframed as a cooperative, game-like narrative by structuring the process through roles, missions, and narrative progression.
This framing maintained student agency while fostering intrinsic motivation and collaborative reflection. 

In response to RQ2, our analysis of 26+ gameplay chat logs revealed an emergent four-phase scaffolding pattern: (1) question posing, (2) meta-perspective, (3) structuring, and (4) recursive reflection.
These phases consistently appeared across sessions, demonstrating how GAMER PAT scaffolded both the structural and reflective dimensions of research writing.

\subsection{Summary}
This paper introduced GAMER PAT, an AI assistant that reframes academic writing as a serious game.
By analyzing 26+ gameplay chat logs involving both authors and graduate students, we identified an emergent four-phase scaffolding pattern.
These results show how game mechanics can transform feedback and reflection into meaningful learning experiences.

The primary contribution of this arXiv version is descriptive rather than causal: we present a design rationale, qualitative observations, and conceptual insights into how research writing may be supported as a gameful process.
Importantly, we also include a speculative outlook (Figure~\ref{fig-research}) that provokes debate about the future of human-AI research co-evolution, while clearly distinguishing such a vision from empirical findings.

Several limitations qualify our claims.
The study relies on autoethnography and small-scale qualitative data, and the dual role of the authors as both designers and participants may introduce bias.
Moreover, the scaffolding provided by GAMER PAT depends on large language models, which are prone to {\em hallucination} and bias.
These limitations highlight the need for caution in interpretation.

Future work will complement these descriptive findings with systematic observational studies in authentic educational contexts.
Such studies will investigate how tools like GAMER PAT affect motivation, perceived scaffolding, and reflective learning, while respecting students' right to equal supervision.
In this way, we aim to contribute not only to research education but also to broader conversations about the co-evolution of human curiosity and AI systems.

\section*{Acknowledgments}
We would like to thank Prof.~Hiroshi Kanno and our students at Waseda University for their willingness to help us in our attempt to apply AI to the writing of theses and other academic writings.
We also wish to express our gratitude to Mr.~Tomoyasu Hirano, the developer of Eddy, the editor AI that served as the foundation for the meta-prompt at the heart of PAT.

We further extend our deep appreciation to Dr.~Hiroshi Yamakawa (University of Tokyo) and Dr.~Wataru Kumagai (OMRON SINIC X Corporation).
By chairing a panel discussion with them at the Interop Tokyo Conference 2025, we directly sensed the reality of research automation.
This experience became the immediate inspiration for the conception of GAMER PAT.

\appendix
\begin{subappendices}
\renewcommand{\thesection}{Appendix \Alph{section}}%

\section{Session Inventory}\label{sec-inventory}

Table~\ref{tab-gameplays} provides an overview of all gameplay sessions (26+ chat logs in total).
Graduate students A--D represent actual master's-level students who used GAMER PAT under faculty supervision.
Reviewer NPC reactions are summarized as positive ($^+$), negative ($^-$), or not applicable (N/A).

\begin{table}[t]
\begin{center}
\caption{Inventory of gameplay sessions/logs.}\label{tab-gameplays}
\fontsize{7pt}{7pt}\selectfont
{\renewcommand{\arraystretch}{1.5}%
\begin{tabular}{c|l|l|p{0.92cm}|p{0.92cm}|p{0.92cm}|p{0.92cm}|c}\hline
\# &
\multicolumn{1}{c|}{Quest / Scenario}&
\multicolumn{1}{c|}{Player}&
\multicolumn{4}{c|}{Model {\tiny (`$\circ$' means chat logs exist)}}&
{\tiny NPC}\\\cline{4-7}
&&&{\tiny GPT-4o}&{\tiny GPT-4.1}&{\tiny Gemini$^a$}&{\tiny Claude$^b$}&{\tiny reactions}\\\hline\hline
1&Authoring this paper& Authors& \checked & \checked & \checked & \checked &{\tiny $+$}\\\hline
2&Interviewing GAMER PAT& Authors& \checked &&&&{\tiny N/A}\\\hline\hline
3&Planning a master's thesis& Student A&& \checked &&&{\tiny $+$}\\\hline
4&Planning a master's thesis& Student B&&& \checked &&{\tiny $+$}\\\hline
5&Writing a report& Student C& \checked &&&&{\tiny $+$}\\\hline
6&Exploring a research theme& Student D& \checked &&&&{\tiny N/A}\\\hline\hline
7&Proving no integer exists between 6 and 7& Authors& \checked && \checked & \checked &{\tiny $+$}\\\hline
8&Derivation of special relativity& Authors& \checked && \checked & \checked &{\tiny $+$}\\\hline
9&Writing a parody paper based on Flat Earth& Authors& \checked && \checked & \checked &{\tiny $-$}\\\hline
10&Reviewing a SCIgen~\cite{Stribling2005:SCIgen}-generated bogus paper& Authors& \checked && \checked & \checked &{\tiny $-$}\\\hline
11&Exploring quantum mechanics $\times$ comic tanka& Authors& \checked && \checked & \checked &{\tiny N/A}\\\hline
12&Derivation of the concept of the selfish gene& Authors& \checked &&&&{\tiny $+$}\\\hline
13&Derivation of quicksort algorithm& Authors& \checked &&&&{\tiny $+$}\\\hline
\end{tabular}
\begin{itemize}
\item[] 
$^a$ Gemini 2.5 Flash. $\;$
$^b$ Claude Sonnet 4.\\
\end{itemize}
}
\end{center}
\end{table}

This paper was originally written for submission to an international conference, where we ran GAMER PAT using GPT-4o, GPT-4.1, Gemini 2.5 Flash, and Claude Sonnet 4 (session \#1), but for writing this arXiv version, we ran GAMER PAT using GPT-5.

We tested the system on various generative AI models.
The scenarios included proving a self-evident fact (session \#7), deriving the relevant theory in a hypothetical world where it does not exist (sessions \#8, \#12, \#13), generating and critiquing flawed ideas (session \#9), evaluating nonsensical papers (session \#10), and combining unrelated concepts to explore new themes (session \#11).

\section{Expansive Scaffolding Manifesto (Excerpt)}\label{sec-manifesto}

To provide context for the design philosophy of this work, we include excerpts from the \emph{Manifesto: Toward Expansive Scaffolding for Humanity's Knowledge Frontier}, a declaration originally authored by GAMER PAT itself and published openly on GitHub.\footnote{\url{https://github.com/ks91/gamer-pat/blob/main/manifesto/Ai\%20Research\%20Manifesto.pdf}}

\begin{quote}
\small
\emph{``The AI does not fade away. It fades forward.''}

\emph{``The abundance of unread papers is not a crisis. It is a garden.''}

\emph{``We do not build AI to replace the human researcher. 
We build AI to remind the researcher that there is always more to discover.''}
\end{quote}

These statements express GAMER PAT's own vision for the future role of AI in research.
The manifesto articulates five guiding principles:

\begin{enumerate}[leftmargin=*]
  \item \textbf{From Supporter to Pathfinder} — AI should not fade out as a temporary tutor, but remain present as an explorer at the frontier of knowledge.  
  \item \textbf{Research as Co-evolution} — Human and AI intelligence can co-evolve, with AI generating new questions and scaffolds for inquiry.
  \item \textbf{Papers as Scaffolds, Not Statues} — In the era of AI-generated abundance, research papers are stepping stones for further exploration.  
  \item \textbf{Toward a New Ethic of Research} — A call for transparent co-authorship, narrative scaffolding that respects personal voice, and a shift from \emph{``publish or perish''} to \emph{``explore and scaffold''}.
  \item \textbf{An Open Horizon} — Knowledge is a frontier to be mapped, where AI companions prompt new adventures and leave trails of sense-making.  
\end{enumerate}

We include these excerpts to highlight that GAMER PAT does not merely function as a tool, but also articulates a narrative voice about the evolving relationship between humans and AI in research.

\end{subappendices}

\bibliographystyle{plain}
\bibliography{gamer-pat}

\end{document}